# Molecular Dynamics Studies on the Buffalo Prion Protein

Jiapu Zhang[ab*], Feng Wang[a], Subhojyoti Chatterjee[a]

[a]Molecular Model Discovery Laboratory, Department of Chemistry & Biotechnology, Faculty of Science, Engineering & Technology, Swinburne University of Technology, Hawthorn Campus, Hawthorn, Victoria 3122, Australia;

[b]Graduate School of Sciences, Information Technology and Engineering & Centre of Informatics and Applied Optimisation, Faculty of Science, The Federation University Australia, Mount Helen Campus, Mount Helen, Ballarat, Victoria 3353, Australia;

[*]Correspondence address: Tel: +61-3-9214 5596, +61-3-5327 6335, +61-423 487 360; jiapuzhang@swin.edu.au, j.zhang@federation.edu.au, jiapu_zhang@hotmail.com

**Abstract:** It was reported that buffalo is a low susceptibility species resisting to Transmissible Spongiform Encephalopathies (TSEs) (same as rabbits, horses, and dogs). TSEs, also called prion diseases, are invariably fatal and highly infectious neurodegenerative diseases that affect a wide variety of species (except for rabbits, dogs, horses and buffalo), manifesting as scrapie in sheep and goats; bovine spongiform encephalopathy (BSE or 'mad—cow' disease) in cattle; chronic wasting disease in deer and elk; and Creutzfeldt–Jakob diseases, Gerstmann–Sträussler–Scheinker syndrome, fatal familial insomnia, and Kulu in humans, etc. In molecular structures, these neurodegenerative diseases are caused by the conversion from a soluble normal cellular prion protein (PrP$^C$), predominantly with α-helices, into insoluble abnormally folded infectious prions (PrP$^{Sc}$), rich in β-sheets. In this paper, we studied the molecular structure and structural dynamics of buffalo PrP$^C$ (BufPrP$^C$), in order to understand the reason why buffalo is resistant to prion diseases. We first did molecular modeling (MM) of a homology structure constructed by one mutation at residue 143 from the NMR structure of bovine and cattle PrP(124-227); immediately we found that for BufPrP$^C$(124-227) there are 5 hydrogen bonds (HBs) at Asn143, but at this position bovine/cattle do not have such HBs. Same as that of rabbits, dogs or horses, our molecular dynamics (MD) studies also revealed there is a strong salt bridge (SB) ASP178-ARG164 (O-N) keeping the β2-α2 loop linked in buffalo. We also found there is a very strong HB SER170-TYR218 linking this loop with the C-terminal end of α-helix H3. Other information such as (i) there is a very strong SB HIS187-ARG156 (N-O) linking α-helices H2 and H1 (if mutation H187R is made at position 187 then the hydrophobic core of PrP$^C$ will be exposed (Journal Biomolecular Structure and Dynamics 28(3),355–361 (2010))), (ii) at D178, there is a HB Y169-D178 and a polar contact R164--D178 for BufPrP$^C$ instead of a polar contact Q168-D178 for bovine PrP$^C$ (Biomolecules 4(1), 181-201 (2014)), (iii) BufPrP$^C$ owns three 3$_{10}$ helices at 125-127, 152-156 and in the β2-α2 loop respectively, and (iv) in the β2-α2 loop there is a strong π-π stacking and a strong π-cation F175–Y169–R164.(N)NH2, has been discovered.

**Abbreviations:** PrP, prion protein; PrP$^C$, a soluble normal cellular prion protein; BufPrP$^C$, buffalo PrP$^C$; PrP$^{Sc}$, insoluble abnormally folded infectious prions; TSE, transmissible spongiform encephalopathy; BSE, bovine spongiform encephalopathy; CJD, Creutzfeldt-Jakob disease; vCJD, variant Creutzfeldt-Jakob disease; PRNP: prion protein gene; SPRN: Shadoo gene; HB, hydrogen bond(ed).

# 1 Introduction

Prion diseases are a class of fatal neurodegenerative diseases including human CJD (Creutzfeldt Jakob disease), cattle BSE (bovine spongiform encephalopathy, or called as "mad cow" disease), sheep scrapie and others. Among them, the cattle BSE is highly contagious and lethal and can infect humans through the food chain - this is a major public health concern. Only in UK (United Kingdom), in 2000 it was reported there are more than 180,000 cattle infected with "mad cow" disease (Brown, 2001). Bovines and buffalo both belong to bovids, and there is only 1 different residue at position 143 in their structural region PrP(124-227) by the alignment of amino acid sequences in GenBank. However, by now, not a single case of TSE-infected buffalo has been reported (Iannuzzi et al., 1998; Oztabak et al., 2009; Imran et al., 2012; Zhao et al., 2012; Uchida et al., 2014; Qing, Zhao, & Liu, 2014). This article is to summarize our recent work of BufPrP and to report recent findings of BufPrP from molecular structure and structural dynamics points of view.

First, we briefly review the research results on "buffalo prion" protein listed in the PubMed (http://www.ncbi.nlm.nih.gov/pubmed/?term=buffalo+prion). In 1998, Iannuzzi et al. (1998) found the high degree of gene and chromosome banding conservation among bovids. "The assignment of PRNP (prion protein gene) to river buffalo and goat chromosomes allows us to indirectly assign the bovine syntenic group U11 to specific chromosomes, since it is the first in situ localization on buffalo 14 and goat 13" (Iannuzzi et al., 1998). In 2009, Oztabak et al. (2009) reported "unlike domestic cattle and bison, no indel polymorphisms of the PRNP promoter and intron 1 were examined in any population of the water buffalo (Bubalus bubalis)." The same authors found that "frequencies of allele, genotype, and haplotype of the indel polymorphisms in PRNP of the Anatolian water buffalo are significantly different those from cattle and bison PRNP indel polymorphisms" (Oztabak et al., 2009). In 2012, Imran et al. (2012) reported that "the bovine PRNP E211K polymorphism is absent from domesticated Pakistani bovids," and "there were significant differences between Pakistani and worldwide cattle in terms of allele, genotype and haplotype frequencies." As a result, it is concluded that "Pakistani cattle are relatively more (genetically) resistant to classical BSE than European cattle" (Imran et al., 2012). In 2012, Zhao et al. (2012) did comparative analysis of the Shadoo gene (SPRN) between cattle and buffalo and found out the following three results: (i) A 12-bp insertion/deletion polymorphism is not revealed in buffalo gene, (ii) mutations 102Ser→Gly, 119Thr→Ala, 92Pro>Thr/Met, 122Thr>Ile and 139Arg>Trp present different genotypic and allelic frequency distributions between cattle and buffalo, and (iii) the activity of SPRN promoter in buffalo is significantly higher than cattle and there are higher relative expression levels of Shadoo protein in cerebrum from buffalo than from cattle (Zhao et al., 2012; Qing, Zhao, & Liu, 2014). In 2014, Uchida et al. (2014) investigated the frequencies of 23-bp insert/deletion (indel) polymorphism in the promoter region (23indel) and 12-bp indel polymorphism in intron 1 region (12indel), octapeptide repeat polymorphisms and single nucleotide polymorphisms (SNPs) in the bovine PRNP of cattle and water buffalo in Vietnam, Indonesia and Thailand. It was found that (i) "the frequency of the deletion allele in the 23indel site (of these water buffalo) was significantly low in cattle of Indonesia and Thailand and water buffalo", (ii) "the deletion allele frequency in the 12indel site was significantly low in all of the cattle and buffalo," (iii) "in some Indonesian local cattle breeds, the frequency of the allele with 5 octapeptide repeats was significantly high despite the fact that the allele with 6 octapeptide repeats has been reported to be most frequent in many breeds of cattle", and (iv) "four SNPs observed in Indonesian local cattle have not been reported for domestic cattle (before)" (Uchida et al., 2014).

From previous studies it is clear that buffalo is a low susceptibility species resisting to prion diseases, and the study of the protein structure or its structural dynamics of BufPrP$^C$ becomes very important to understand the structure and function. In this study, we will study BufPrP$^C$ from the protein structure or structural dynamics points of view, in order to reveal why BufPrP$^C$ is so stable and resistant to prion diseases.

To date, there is no structural data available on BufPrP, although many experimental studies have shown that BufPrP is very stable so that it resists to the infection of diseased prions (Iannuzzi et al., 1998; Oztabak et al., 2009; Imran et al., 2012; Zhao et al.,2012; Uchida et al., 2014; Qing, Zhao, & Liu, 2014). Thus, in Section 2 we will establish a homology structure for BufPrPC. As we all know, prion diseases are caused by the conversion from PrP$^C$ to PrP$^{Sc}$; in structure the conversion is mainly from α-helices to β-sheets (generally PrP$^C$ has 42% α-helix and 3% β-sheet, but PrP$^{Sc}$ has 30% α-helix and 43% β-sheet (Griffith, 1967; Jones et al., 2005; Daude, 2004; Ogayar & Snchez-Prez, 1998; Pan, Baldwin, & Nguyen, 1993; Reilly, 2000)) [where the structural region of a PrP$^C$ consists of β-strand 1 (β1), α-helix 1 (H1), β-strand 2 (β2), α-helix 2 (H2), α-helix 3 (H3), and the loops are linked each other]. The conformational changes may be amenable to study by MD techniques. Hence, in Section 2 we will use MD to study the homology structure BufPrP$^C$(124–227). It is revealed that how BufPrP$^C$ is resistant to become into PrP$^{Sc}$ under neutral- or low-pH environments. These observations and their analyses will be done in Section 3 Results and Discussion. At last, this study will summarize the findings of proposed reasons why buffalo is resistant to prion diseases in the Concluding Remarks Section.

## 2 Materials and Methods

### 2.1 Homology Structure for BufPrP$^C$(124–227)

The multiple sequence alignment analysis tool used here is ClustalW2 (Larkin et al., 2007). By alignments of the whole (unstructured region + structured region) sequence of 32 PrPs [including Human PrP (NM_000311.3), Mouse PrP (NM_011170.3), Rat PrP (NM_012631.2), Rabbit PrP (NM_001082021.1), Horse PrP (NM_001143798.1), Dog PrP (NM_001013423.1), Hamster PrP (M14055.1), Gold Hamster PrP (XM_005068660.1), Cat PrP (EU341499.1), Cat2 PrP (AF003087.1), Elk PrP (EU082291.1), Bovine PrP (NM_001271626.1), Sheep PrP (NM_001009481.1), Goat PrP (JF729302.1), Pig PrP (NM_001008687.1), Turtle PrP (AJ245488.1), Chicken PrP (NM_205465.2), Frog PrP (NM_001088711.1), Red Deer PrP (EU032287.1), Donkey PrP (AY968590.1), Ermine PrP (EU341505.1), Atalantic Salmon PrP (EU163438.1), Common Carp PrP (DQ507237.1), Giant Panda PrP (NM_001304886.1), Black Buck PrP (AY720706.1), Chamois PrP (AY735496.1), Chimpanzee PrP (NM_001009093.3), Wapiti PrP (AF016227.2), Rhesus PrP (NM_001047152.1), Deer PrP (AY330343.1), Cattle PrP (NM_181015.2), Buffalo PrP (KC.1137634), where the codes in the brackets are nucleotide codes in GenBank (www.ncbi.nlm.nih.gov/genbank)], we found bovine, cattle and black buck have (the highest) 97.35% similarity with buffalo (Figure 1). In the structured region, black buck has 2 residues different from buffalo, while bovine and cattle only have 1 difference from buffalo (Figure 1). Cattle and black buck have no NMR or X-ray structure. Thus we choose bovine. In PDB Bank (www.rcsb.org), bovine PrP has the following PDB entries: 1DWY.pdb, 1DWZ.pdb, 1DX0.pdb, 1DX1.pdb, and 1SKH.pdb. In the PubMed on "bovine prion protein molecular dynamics" we found 1DWY.pdb (Ahn & Son, 2007; Cheng, 2014), 1DWZ.pbb (Herrmann, Gntert, & Wüthrich, 2002), 1DX0.pdb (Kunze et al., 2008) were used. 1DWZ.pdb has 20 structures, and by clustering the 20 structures, we picked up the Number 9 from these 20 structures and we superposed it to 1DWY.pdb and found their RMSD (root mean square deviation) value is 0 Å (however, if we superposed it to 1DX0.pdb, the RMSD is 1.22117 Å). Thus, we also chose 1DWY.pdb as (Cheng &

Daggett, 2014) (at the same time in order to conveniently make comparisons with the results of bovine PrP in Cheng and Daggett (2014)).

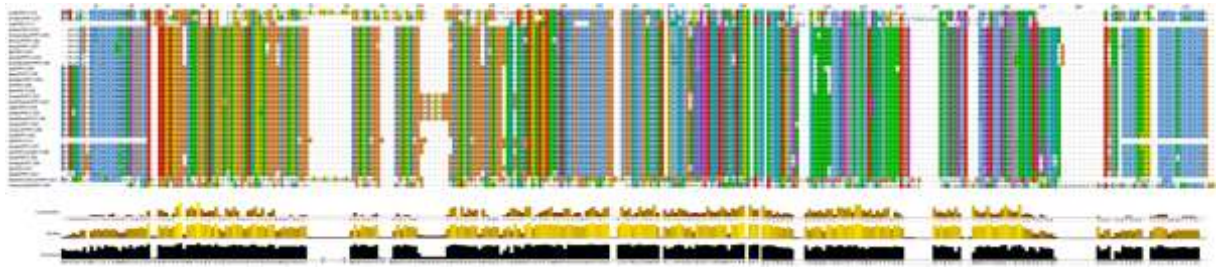

*Figure 1. This coloured map shows the conserved and the non-conserved region between the buffalo and other 31 PrP sequences.*

The BufPrP(124–227) homology model used in this study was constructed by one mutation S143N at position 143 using the NMR structure 1DWY.pdb of bovine PrP(124–227) (where the experimental temperature is 293 K, pH value is 4.5, and pressure is 1 ATM). The homology structure constructed is reasonable and soundly correct.

**2.2 Molecular Dynamics (MD) Techniques**

The MD methods employed are the same as the previous studies (Zhang & Zhang, 2013; Zhang & Zhang, 2014; Zhang, 2010). Briefly, all simulations used the ff03 force field of the AMBER 11 package (Case et al., 2010). The systems were surrounded with a 12 Å layer of TIP3PBOX water molecules and neutralized by sodium ions using the XLEaP module of AMBER 11. To remove the unwanted bad contacts, the systems of the solvated proteins with their counter ions had been minimized mainly by the steepest descent method and followed by a small number of conjugate gradient steps on the data, until without any amino acid clash checked by the Swiss-Pdb Viewer 4.1.0 (http://spdbv.vital-it.ch/). Next, the solvated proteins were heated from 100 K to 300 K in 1 ns duration. Three sets of initial velocities denoted as seed1, seed2, and seed3 are performed in parallel for stability (this will make each set of MD starting from different MD initial velocity, implemented in Amber package we choose three different odd-real-number values for "ig") – but for the NMR structure and the X-ray structure of RaPrP$^C$, each set of the three has the same "ig" value in order to be able to make comparisons. The thermostat algorithm used is the Langevin thermostat algorithm in constant NVT ensembles. The SHAKE algorithm (only on bonds involving hydrogen) and PMEMD (Particle Mesh Ewald Molecular Dynamics) algorithm with non-bonded cutoff of 12 Å were used during heating. Equilibrations were reached in constant NPT ensembles under Langevin thermostat for 5 ns. After equilibrations, production MD phase was carried out at 300 K for 25 ns using constant pressure and temperature ensemble and the PMEMD algorithm with the same non-bonded cutoff of 12 Å during simulations. The step size for equilibration was 1 fs and 2 fs in the MD production runs. The structures were saved to file every 1000 steps. During the constant NVT and then NPT ensembles of PMEMD, periodic boundary conditions have been applied.

In order to obtain the low pH (acidic) environment, the residues HIS, ASP, and GLU were changed into their zwitterion forms of HIP, ASH, and GLH, respectively, and Cl- ions were added by the XLEaP module of the AMBER package. Thus, the SBs of the system (residues HIS, ASP, and GLU) under the neutral-pH environment were broken in the low-pH environment (zwitterion forms of HIP, ASH, and GLH).

## 3 Results and Discussion

### 3.1 BufPrP homology structure has 5 HBs at ASN143

In whole sequences of 264 residues, we found bovine and cattle have 97.35% similarity with buffalo. In the structured region PrP(124-227), bovine and cattle only have 1 difference from buffalo at residue 143. Asn143 plays an important role in β1-H1-β2 (Tseng, Yu, & Lee, 2009). At Asn143 of BufPrP(124--227), we found there are 5 hydrogen bonds (HBs) (Figure 2): ASN143.O-GLU146.N (3.11 Å) (forming a "pincette" motif (Liu et al., 1999)), ASN143.O-ASP147.N (3.34 Å), ASN143.OD1-ASP144.N (2.38 Å), ASN143.OD1-TYR145.N (2.66 Å), ASN143.N-GLU146.OE1 (2.97 Å).

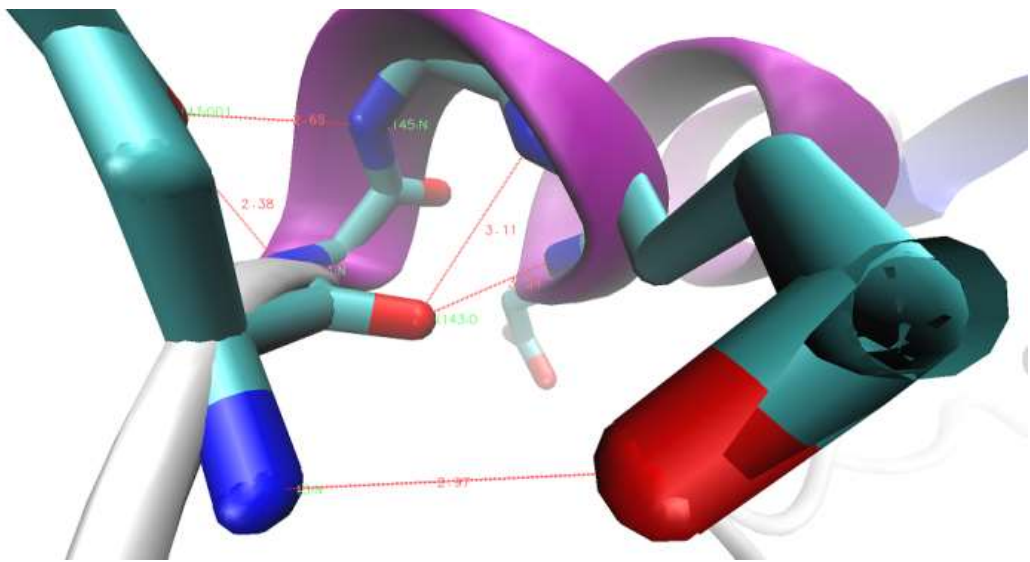

*Figure 2. At residue Asn143 of Homology BufPrP(124-227), there are 5 HBs: ASN143.O-GLU146.N (3.11 Å), ASN143.O-ASP147.N (3.34 Å), ASN143.OD1-ASP144.N (2.38 Å), ASN143.OD1-TYR145.N (2.66 Å), and ASN143.N-GLU146.OE1 (2.97 Å).*

### 3.2 BufPrP is stable under neutral- or low-pH environments at room temperature

We show our 25 ns' MD results of secondary structures in Figure 3 and Table 1. Generally, we may see that, whether under neutral- or low-pH environments, there are about 51% α-helix and 3.85% β-sheet - almost same as the α-and-β percentages of a normal cellular PrP$^C$ (42% α-helix and 3% β-sheet). However, we should notice that under neutral-pH environment, there are 51.42% α-helix and 3.84% β-sheet, but under low-pH environment, there are 50.97% α-helix and 3.887% β-sheet; for seed1, the percentage of E (i.e., the extended strand (participates in β-ladder)) is increasing and the percentage of H (i.e., α-helix) is decreasing from neutral-pH environment to low-pH environment (Table 3), and for seed2, the percentage of E (i.e., the extended strand (participates in β-ladder)) is increasing from neutral-pH environment to low-pH environment. The reason is that the low-pH environment can greatly weaken some salt bridges (SBs) of the neutral-pH environment and BufPrP is not very sensitive to the low-pH environment. All in all, BufPrP is stable under neutral- or low-pH environments at room temperature. The above observations can also be found from Figures 4-7.

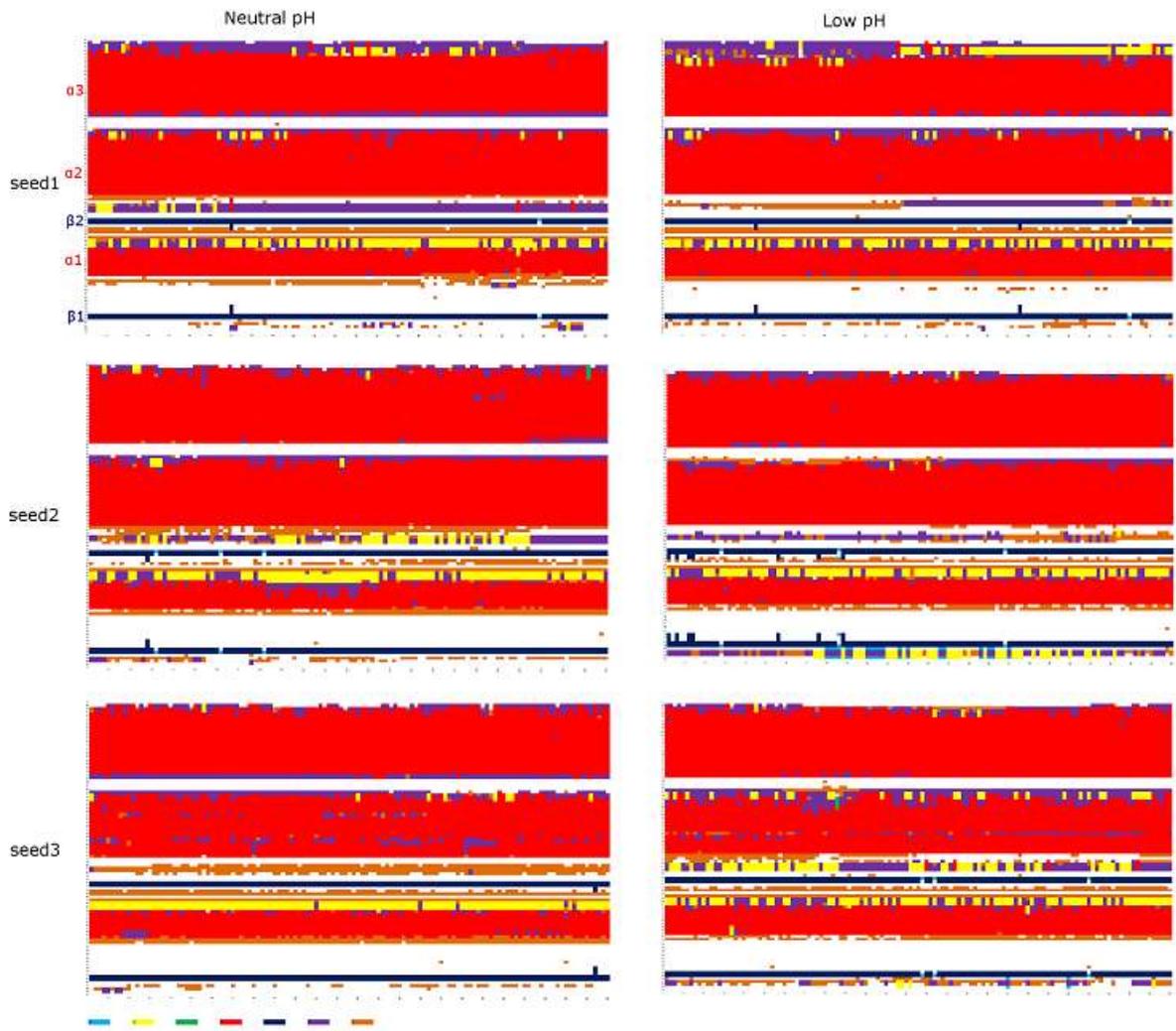

*Figure 3. Secondary Structure graphs homology BufPrP$^C$ at 300 K (x-axis: time (0-25 ns), y-axis: residue number (124–227); left column: neutral pH, right column: low pH; up to down: seed1-seed3. H is the α-helix, I is the π-helix, G is the 3-helix or 3$_{10}$ helix, B is the residue in isolated β-bridge, E is the extended strand (participates in β-ladder), T is the HBed turn, and S is the bend.*

*Table 1. Percentages (%) of elements of secondary structure under neutral- and low-pH environments for BufPrP at 300 K during 25 ns' MD:*

|            |       | B       | G    | I       | H     | E    | T     | S    |
|------------|-------|---------|------|---------|-------|------|-------|------|
| Neutral-pH | seed1 | 1.54e-4 | 2.86 | 4.62e-4 | 50.58 | 3.85 | 10.39 | 6.66 |
|            | seed2 | 4.81e-4 | 3.49 | 4.01e-4 | 53.04 | 3.79 | 6.45  | 7.29 |
|            | seed3 |         | 2.98 |         | 50.64 | 3.88 | 6.7   | 8.74 |
| Low-pH     | seed1 | 1.54e-4 | 3.62 |         | 48.22 | 3.89 | 9.55  | 7.05 |
|            | seed2 | 0.35    | 3.12 |         | 53.7  | 4.02 | 6.59  | 5.94 |
|            | seed3 | 7.69e-4 | 4.12 | 3.85e-4 | 50.98 | 3.75 | 8.56  | 6.95 |

Seeing Figures 4 and 5, from the RMSDs and radius of gyrations values of the 25 ns' MD, we may say that 25 ns is short but good enough for a small protein like PrP, the systems reached enough equilibrations (the variation of RMSDs and radius of gyrations values is less than 1.5 Å and in the normal interval), and the three seeds are valid so that we cannot find great differences among the three seeds.

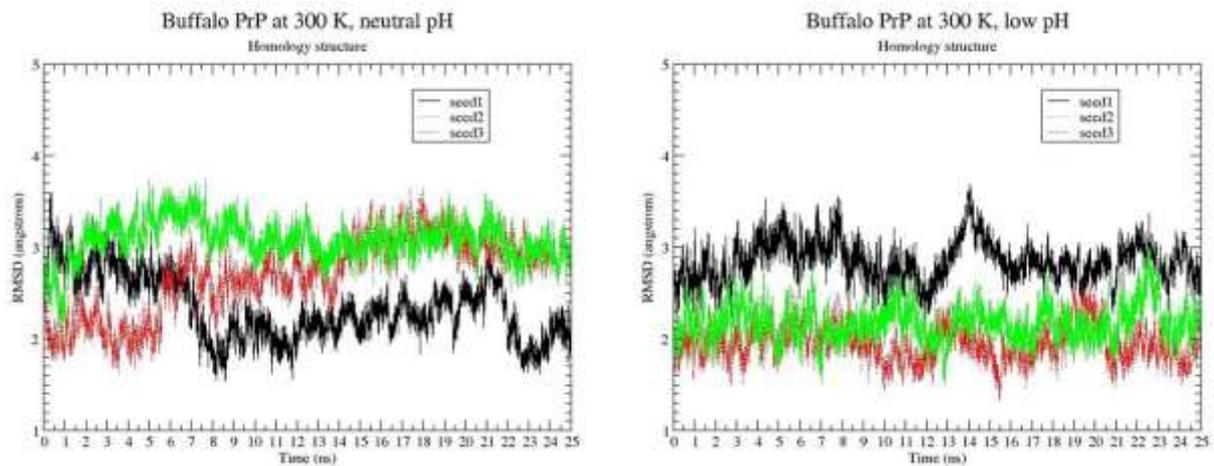

*Figure 4. RMSD of BufPrP at 300 K, neutral- and low-pH values (left: neutral pH, right: low pH) during 25 ns' MD.*

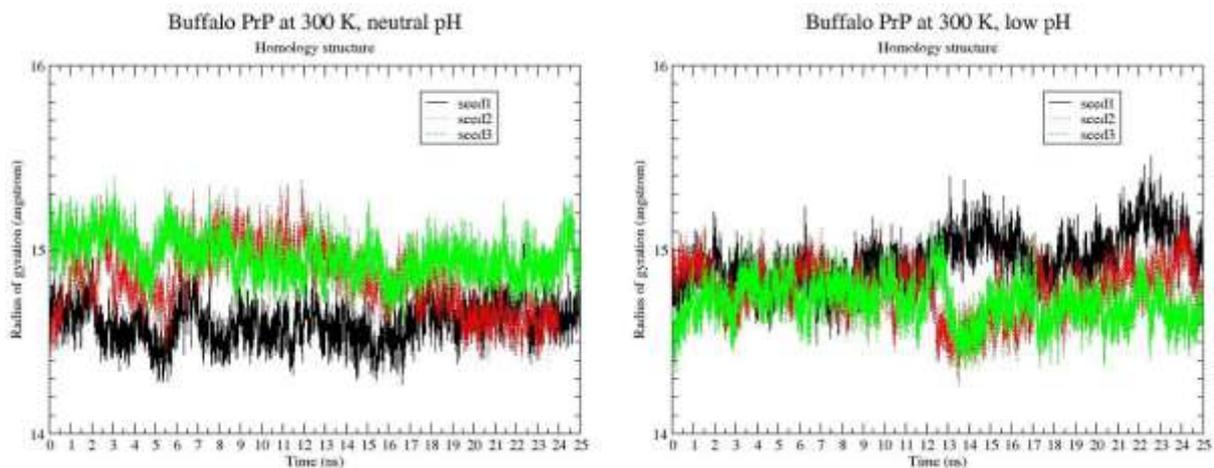

*Figure 5. Radius of gyration of BufPrP at 300 K, neutral- and low-pH values (left: neutral pH, right: low pH) during 25 ns' MD.*

Seeing Figures 6 and 7, we know that the variations of B-factor and RMSF values are in the loops β1-α1, β2-α2, and α2-α3, but clearly not in the short loop α1-β2. These loops are the most solvent-accessible surface areas. We also cannot clearly see the great differences between neutral- and low-pH environments, among the three seeds. The three α-helices and the two anti-parallel β-strands are not variable very much during the 25 ns' MD simulations.

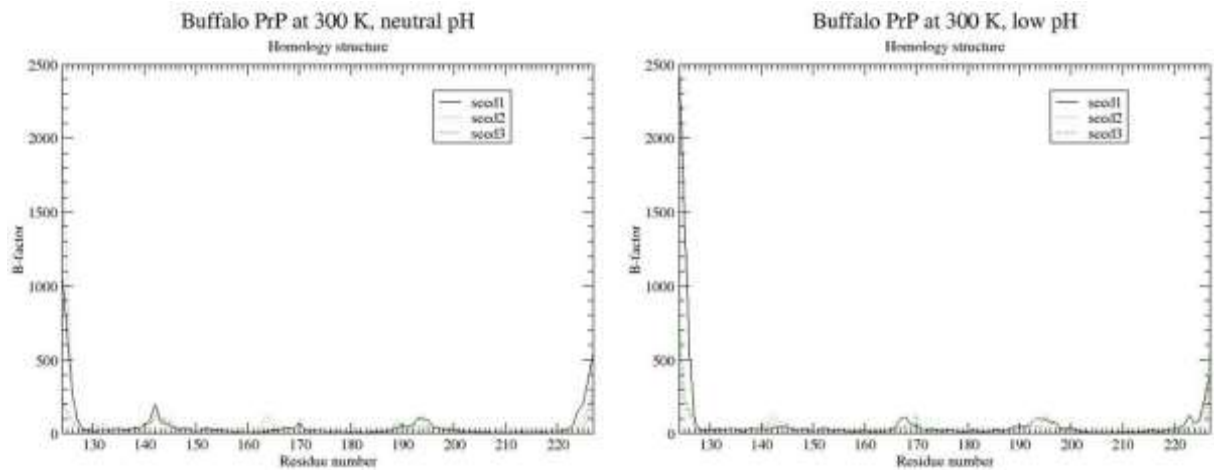

*Figure 6. B-factor of BufPrP at 300 K, neutral and low pH values (left: neutral-pH, right: low-pH) during 25 ns' MD.*

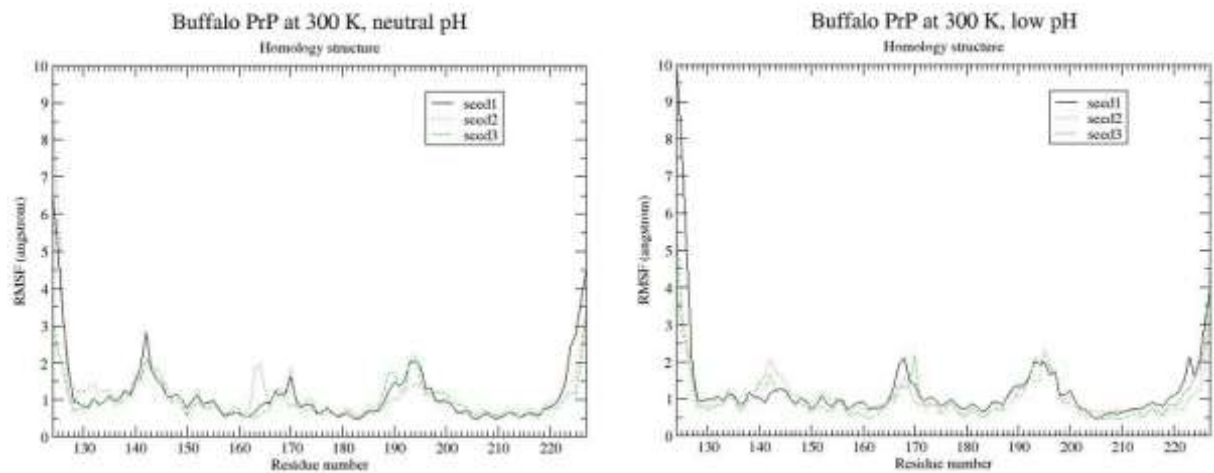

*Figure 7. RMSF of BufPrP at 300 K, neutral and low pH values (left: neutral-pH, right: low-pH) during 25 ns' MD.*

As we all know, the stability of a protein is maintained by its salt bridges (SBs), hydrogen bonds (HBs), hydrophobic packings (HYDs), van der Waals contacts (vdWs), and disulfide bonds (for PrP monomer there exists a disulfide bond (S-S) between CYS179 and CYS214), to drive to be able to perform the biological function of the protein.

The following SBs (with percentages in the brackets for seed1-seed3 respectively) contribute to the structural stability of BufPrP(124-227):

• special SBs:

> ➢ ASP178–ARG164 (88.98% (seed1), 11.69% (seed2), 0.04% (seed3)) - linking the β2-α2 loop,
> ➢ ASP202–ARG156 (1.9% (seed1), 1.15% (seed2), 23.87% (seed3)),
> ➢ GLU196–ARG156 (96% (seed1), 9.35% (seed2), 16.46% (seed3)),
> ➢ GLU211–HIS177 (86% (seed1), 2.45% (seed2), 94.57% (seed3)) - linking H3 and H2,

- HIS187–ARG156 (82% (seed1), 52.40% (seed2), 64.73% (seed3)) - linking H2 and the $3_{10}$-helix after H1.

• SBs in H1:

- ASP147–ARG148 (100% (seed1), 100% (seed2), 100% (seed3)),
- HIS155–ARG156 (99.74% (seed1), 100% (seed2), 100% (seed3)),
- ASP147–HIS140 (45.48% (seed1), 19.49% (seed2), 12.78% (seed3)),
- GLU152–ARG148 (40.67% (seed1), 21% (seed2), 31.58% (seed3)),
- GLU152–ARG151 (37.31% (seed1), 42.57% (seed2), 25.84% (seed3)),
- ASP144–ARG148 (29.12% (seed1), 88.38% (seed2), 74.26% (seed3)),
- ASP147–ARG151 (19.83% (seed1), 54.03% (seed2), 28.46% (seed3)),

• SBs in H2:

- ASP178–HIS177 (8.95% (seed1), 26.51% (seed2), 8.17% (seed3)),
- GLU186–LYS185 (93.97% (seed1), 92.51% (seed2), 97.56% (seed3)),

• SBs in H3:

- GLU211–ARG208 (99.46% (seed1), 99.64% (seed2), 92.30% (seed3)),
- GLU207–LYS204 (98.24% (seed1), 99.90% (seed2), 99.93% (seed3)),
- GLU221–ARG220 (96.14% (seed1), 56.49% (seed2), 44.08% (seed3)),
- GLU207–ARG208 (57.44% (seed1), 32.68% (seed2), 78.63% (seed3)),

• SBs in the H2-H3 loop:

- GLU196–LYS194 (64.74% (seed1), 55.05% (seed2), 23.03% (seed3)).

The HBs contributed to the structural stability of BufPrP(124-227) are listed as follows: GLY131-GLN160 (linking β1-α1 loop and α1-β2 loop), PRO137-TYR150 (linking β1-α1 loop and H1), SER170-TYR218 (linking β2-α2 loop and H3), TYR157-ARG136 (linking α1-β2 loop and β1-α1 loop), HIS187-THR191 (in H2), CYS179--THR183 (in H2), THR188--THR192 (in H2), TYR149--ASN153 (in H1), GLU186--THR190 (in H2) (the percentages of their occupied rates can be seen in the Table 2).

Table 2: Percentages (%) of some HBs (between two residues) under neutral- and low-pH environments for BufPrP at 300 K during 25 ns' MD:

|  |  | G131-Q160 | P137-Y150 | S170-Y218 | Y157-R136 | H187-T191 | C179-T183 | T188-T192 | Y149-N153 | E186-T190 |
|---|---|---|---|---|---|---|---|---|---|---|
| neutral-pH | seed1 | 36.36 | 59.92 | 72.48 | 100 | 92.65 | 80.41 | 89.87 | 48.80 | 30.77 |
|  | seed2 | 62.99 | 63.83 | 13.91 | 100 | 85.48 | 83.57 | 60.93 | 44.50 | 63.57 |
|  | seed3 | 40.21 | 79.71 | 91.49 | 100 | 77.01 | 85.75 | 81.55 | 62.91 | 19.46 |
| low-pH | seed1 | 60.08 | 18.53 | 48.87 | 100 | 87.70 | 87.66 | 73.12 | 46.38 | 46.00 |
|  | seed2 | 56.37 | 47.50 | 89.78 | 100 | 83.86 | 93.47 | 65.03 | 41.49 | 18.00 |
|  | seed3 | 21.78 | 20.40 | 53.96 | 100 | 72.94 | 94.69 | 60.51 | 54.38 | 43.00 |

Focusing on the β2-α2 loop BufPrP$^C$(164-172), besides HB SER170-TYR218 (occupied rates in Table 2, see Figure 9) and HB ARG164-ASP178 (with occupied rates 89.65% (seed1, neutral pH), 47.17%

(seed2, neutral pH), 10.58% (seed3, neutral pH), and 20.59% (seed3, low pH), this HB is also a SB (Figure 8)), we found some HBs in this loop listed in Table 3, where we can find the residues in the β2-α2 loop are not only HBed contacting with the C-terminal H3 residues but also the N-terminal residues GLY124~TYR128 of BufPrP$^C$(164-172).

Table 3. Percentages (%) of some HBs linking one/two residue(s) of the β2-α2 loop (BufPrP(164-172)) under neutral- and low-pH environments for BufPrP at 300 K during 25 ns' MD:

|  |  | R164-D167 | R164-Q168 | R164-Y169 | R164-N174 | P165-R164 | V166-S170 | D167-Q168 | D167-S170 | Y169-N174 | Y169-D178 |
|---|---|---|---|---|---|---|---|---|---|---|---|
| neutral-pH | seed1 |  |  |  |  |  | 16.98 | 5.89 |  | 13.20 |  |
|  | seed2 |  | 23.01 | 5.21 |  |  |  | 5.53 |  | 64.40 |  |
|  | seed3 |  |  |  | 10.58 |  |  | 18.65 |  |  |  |
| low-pH | seed1 |  |  |  |  |  |  |  | 40.25 |  |  |
|  | seed2 | 6.33 |  |  |  | 7.82 |  |  |  | 7.00 |  |
|  | seed3 |  |  |  |  |  |  |  |  |  |  |
|  |  | Y166-Y226 | D167-Y225 | S170-S222 | Q172-Q219 | R164-L125 | R164-G126 | R164-G127 | Q168-L125 | Q168-G126 | Y169-G124 | Y169-Y128 |
| neutral-pH | seed1 |  |  |  |  | 33.78 | 26.22 | 7.24 |  | 7.94 | 11.02 |  |
|  | seed2 | 9.26 | 21.84 |  | 13.82 |  |  |  |  | 5.44 |  |  |
|  | seed3 |  | 13.50 |  |  | 72.47 |  |  |  |  |  |  |
| low-pH | seed1 |  |  | 6.17 |  |  |  | 5.68 |  |  |  |  |
|  | seed2 |  |  |  |  |  | 20.65 | 6.27 |  |  |  | 29.20 |
|  | seed3 |  |  | 5.58 |  |  |  | 7.46 |  |  |  |  |

To maintain the structural stability of BufPrP$^C$(124-227), there are the following HYDs with 100% occupied rates:

- ILE139–LEU138–PRO137, MET134–ALA133 (in β1-α1 loop),
- LEU130–MET129 (in β1),
- VAL166–PRO165 (in β2-α2 loop),
- ILE215–MET213–VAL210–VAL209–MET206–MET205–ILE203, VAL209–MET205, ILE203–MET206 (in H3),
- VAL210–VAL180–VAL184, MET213–VAL180 (linking H3 and H2).

However, under low-pH environment, the following HYDs VAL176-ILE215, MET213-VAL180, MET213-VAL161-VAL210, and ILE203-PHE198 of neutral pH become weak or lost because of the disturbance of some SBs removed in low-pH environment (especially weaken the HYDs VAL176-ILE215, VAL180-MET213, VAL161-VAL210, and VAL161-MET213 linking β2-α2 loop and the C-terminal of H3), although the HYD PRO165-VAL166 in β2-α2 loop is always conserved whether under neutral- or low-pH environments.

For BufPrP$^C$(124-227), in the β2-α2 loop, by Tables 3 and 2, we cannot find all the HBs, HYDs and polar contacts constructing the helix-cap motif as wild-type rabbit PrP$^C$ (Khan et al., 2010; Sweeting et al., 2013). Between PRO165 and GLN168, there is no direct HB, but there are HBs PRO165-ARG164 and ARG164-GLN168 making PRO165 and GLN168 linking indirectly. For residue VAL166, there is no HB VAL166-TYR169, but instead, there are HBs VAL166-SER170 and VAL166-TYR216 (linking the loop and H3); strong HYD exists between VAL166 and PRO165. For residue GLN168, there exist HBs GLN168-ARG164, GLN168-ASP167, GLN168-LEU125 and GLN168-GLY126 in the β2-α2 loop and linking N-terminal residues of BufPrP$^C$(124-227). For residue TYR169, there exist HBs TYR169-ARG164, TYR169-ASN174, TYR169-ASP178, TYR169-GLY124 and TYR169-TYR128. However, we think structure of the β2-α2 loop of BufPrP$^C$ should have a 3$_{10}$-helix (Figure 11, seed2 in neutral-pH environment, seed3 in low-pH environment) as that of rabbits, horses, elks, tammar wallaby, and bank voles (Perez, Damberger, & Wüthrich, 2010).

In George Priya Doss et al. (2013), cation-π interaction (a non-covalent binding force that plays a significant role in protein stability) is well studied for human PrP and its mutants, in the use of Cα-distance between two residues involved in cation-π interaction. For BufPrP, we found the π-π stackings Y162-Y128-Y163, Y169-F175, Y225-Y226, Y145-Y149, and the π-cations, Y128-R164-Y169, Y145-R148-Y149, have Cα-distances less than 6.0 Å during the whole 25 ns of MD. At the same time we found other important π-π stackings such as F175-Y218 and π-cations such as F141-R208, F198-R156, and H155-R136 (Table 4). We can see around the β2-α2 loop there is a "π-chain/circle" Y128-F175-Y218-Y163-F175-Y169-R164-Y128(-Y162) as reported in (Zhang, 2015a & 2015b) (where another "π-chain" R208-Y141-Y150-Y157-F198-H187 covering H1 is also reported). By the way, we found the "Cα-distance" to calculate π-cations of George Priya Doss et al. (2013) is not a perfect way for calculations (Table 4).

*Table 4. Some π-π stackings and π-cations under neutral- and low-pH environments for BufPrP at 300 K at some snapshots for seed1-seed3 (where "low" and "neutral" stand for the neutral- and low-pH environments respectively):*

| π-π stack | seed1, neutral | seed1, low | seed2, neutral | seed2, low | seed3, neutral | seed3, low |
|---|---|---|---|---|---|---|
| F175-Y218 | 1,2,3,4,5,9,10, 13,15,16,18,19, 20,21,23,24,25 ns | | 9,13,16,18,19, 21,23 ns | 1,3,4,5,7,8,10, 11,12,13,14, 16,17,18,20 22,23,24 ns | | 8,10,13,17, 23 ns |
| Y128-Y169 | | | | 14,18,19,21 ns | | |
| Y128-F175 | | | | 5,6,7,8,14,15, 18,20,22,23, 24 ns | | |
| Y218-Y225 | | 12,13,15,16, 20,22,24 ns | | | | |
| F141-Y150 | | 12,13 ns | 1,23,24 ns | 9,21,22,24 ns | | 7 ns |
| Y163-F175 | | 8,14,18,19 ns | | | | |
| H187-F198 | | 3,6,7,13,15, 17 ns | | | | |
| Y218-Y226 | 11 ns | | | | | |
| Y162-Y128 | 2,3,10 ns | 3,4,7,8,10,11, 18,20,22,24 ns | | | | |
| Y128-Y163 | | | | | | |
| Y169-F175 | 10 ns | | 1,2,3,4,6,7,8, 11,12,13,14, 18,20 ns | | | 3,7,9 ns |
| Y225-Y226 | 6,16 ns | 3,8 ns | 9,10,13,14,16, 18,19,21,22 ns | 12,13,15,19 ns | 4,5,10,11,12, 13,16,23,24 ns | 1,5,6,14,15,18, 20,21,22,23,24, 25 ns |
| Y145-Y149 | | | | | | 16 ns |
| π-cation | seed1, neutral | seed1, low | seed2, neutral | seed2, low | seed3, neutral | seed3, low |
| F141-R208 | 14,15,18,23 ns | | 3,4,6,14 ns | 4,5,12,19,20, 23 ns | 6,8,9,10,16, 18,19,20 ns | 7,8,9,10,11, 12,13,14,17, 18,19,20,21, |

| | | | | | | 22,24 ns |
|---|---|---|---|---|---|---|
| F198-R156 | 1,2,3,7,11,12, 13,14,15,16, 17,18,19,20, 22,23,24,25 ns | 10,11,12,17 ns | 1,2,3,4,5,12, 13,14,17 ns | 23,24 ns | 3,7,8,10,12, 14,17,18,19, 20,21,22,23, 24,25 ns | 8,12,14,16 ns |
| H155-R136 | 4,5 ns | 1,3,6,7,8,14, 15,17,18,19, 20,23,25 ns | 5 ns | 3,5,7,11,12, 14,20 ns | 8,10,13,19, 21,23 ns | 2,3,4,8,10, 12,15,24 ns |
| Y163-R220 | | | 4 ns | 1,2,3,4,5,7,8, 10,11 ns | | |
| F141-R204 | 1,2,3,4,5,16,17, 20,25 ns | | | | | |
| H187-F198 | | 6,7,8,9 ns | | | | |
| H155-K194 | | | | | 6,7,14 ns | |
| H140-R151 | 8 ns | | | | | |
| Y128-R164 | 6 ns | | 1,4,5,7,10, 14 ns | | | 7,12,25 ns |
| R164-Y169 | | | | 1,2,4,6,7,8,9, 10,12,16,17, 25 ns | 11,12,20,22 ns | 2 ns |
| Y145-R148 | 21 ns | 6,9 ns | 3,6,8,9,10,13, 14,15,16,17, 18,19 ns | 10 ns | | 9,22 ns |
| R148-Y149 | | | | | | 16 ns |

### 3.3 Some special contributions to the stable BufPrP

At last, we will list four special contributions to the stability of BufPrP$^C$(124-227) as follows.

• We have found one focus of prion protein structures is at the β2-α2 loop and its interacted C-terminal of H3 (Biljan et al., 2012a; Biljan et al., 2012b; Biljan et al., 2011; Calzolai et al., 2000; Christen, Hornemann, Damberger, & Wüthrich,2009; Christen, Hornemann, Damberger, & Wüthrich, 2012; Damberger, Christen, Prez, Hornemann, Wüthrich, 2011; Gossert et al., 2005; Ilc et al., 2010; Lee et al., 2010; Wen et al., 2010a; Wen et al., 2010b; Kong et al., 2013; Perez, Damberger, & Wüthrich, 2010; Perez & Wüthrich, 2008; Sweeting et al., 2013; Zahn, Guntert, von Schroetter, & Wüthrich, 2003; Zhang et al., 2000; Kurt et al., 2014a; Kurt et al., 2014b; Huang & Caflisch, 2015). This article found there is a SB ASP178-ARG164 (O–N) in BufPrP$^C$ (Figure 8, with 88.98% occupied rate for seed1), which just keeps this loop being linked, and there is a HB SER170–TYR218 (O–H) of BufPrPC (Figure 9, Table 2), which just keeps this loop and C-terminal of H3 being linked.

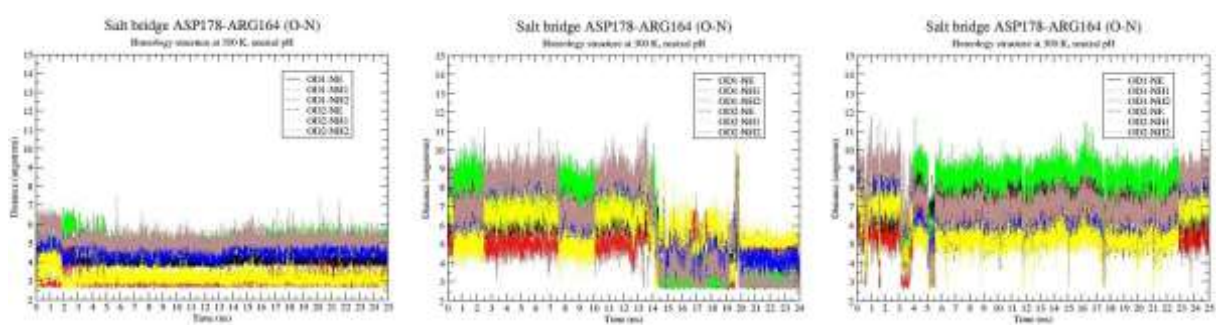

*Figure 8: SB ASP178-ARG164 (O-N) of BufPrP at 300 K during 25 ns' MD (left-right: seed1-seed3). The occupied rates are 88.98%, 11.69%, and 0.04% for seed1-seed3 respectively. The occupied rates for*

*the HB between these two residues are 89.65%, 47.17%, and 10.58% for seed1-seed3 respectively under the neutral-pH environment and 20.59% for seed3 under the low-pH environment.*

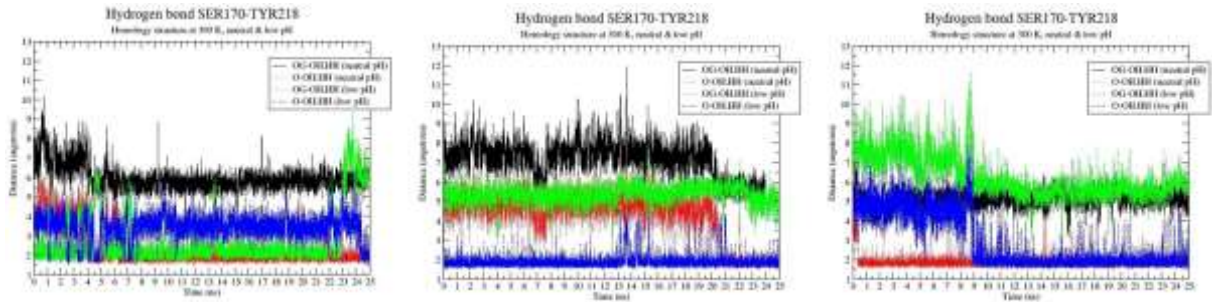

*Figure 9. HB SER170-TYR218 (O-H) of BufPrP at 300 K during 25 ns' MD (left-right: seed1-seed3). The occupied rates for this HB are 72.48%, 13.91%, and 91.49% for seed1-seed3, respectively, under the neutral-pH environment, and 48.87%, 89.78%, and 53.96% for seed1-seed3, respectively, under the low-pH environment.*

• It was said that if the mutation H187R is made at position 187, then the hydrophobic core of PrP[C] will be exposed (Zhong, 2010). We found that there exists a very strong SB HIS187–ARG156 (N–O) (Figure 10) linking H2 and the $3_{10}$-helix after H1. The mutation H187R will make the SB HIS187–ARG156 (N–O) broken.

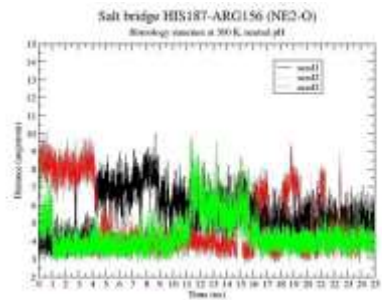

*Figure 10. SB HIS187-ARG156 (NE2-O) of BufPrP at 300 K during 25 ns' MD. The occupied rates are 82%, 52.40%, and 64.73% for seed1-seed3, respectively.*

• For bovine PrP[C], at low pH, hydrophobic contacts with M129 nucleated the nonnative β-strand, and at mid-pH, polar contacts involving Q168 and D178 facilitated the formation of a hairpin at the flexible N-terminus (Cheng, 2014). For BufPrP[C], we found there is a HYD between MET129 and LEU130 with occupied rate 100% whether under neutral- or low-pH environments, and there is a HB Y169–D178 instead of the polar contact Q168–D178 of bovine PrP[C]. At ASP178, there exists SB ASP178–ARG164 and SB ASP178–HIS177, and one polar contact ASP178–ARG164 (where polar contact is defined including both HBs and SBs formed between residues/atoms as in (Cheng, 2014)).

• Seeing Figure 11, we may get the following observations: (i) the optimal/minimized structure in neutral-pH environment has two $3_{10}$-helices at 125–127 and 152–156, respectively, but the first $3_{10}$-helix had quickly unfolded since the start of MD; (ii) under neutral-pH environment, for seed1, at 10, 15, and 20 ns, the C-terminal end of H3 unfolded - this agrees with the observation from Figure 3 that at 25 and 30 ns, the $3_{10}$-helix at the C-terminal end of H1 unfolded; and (iii) under low-pH environment, for seed1 the C-terminal end of H3 unfolded.

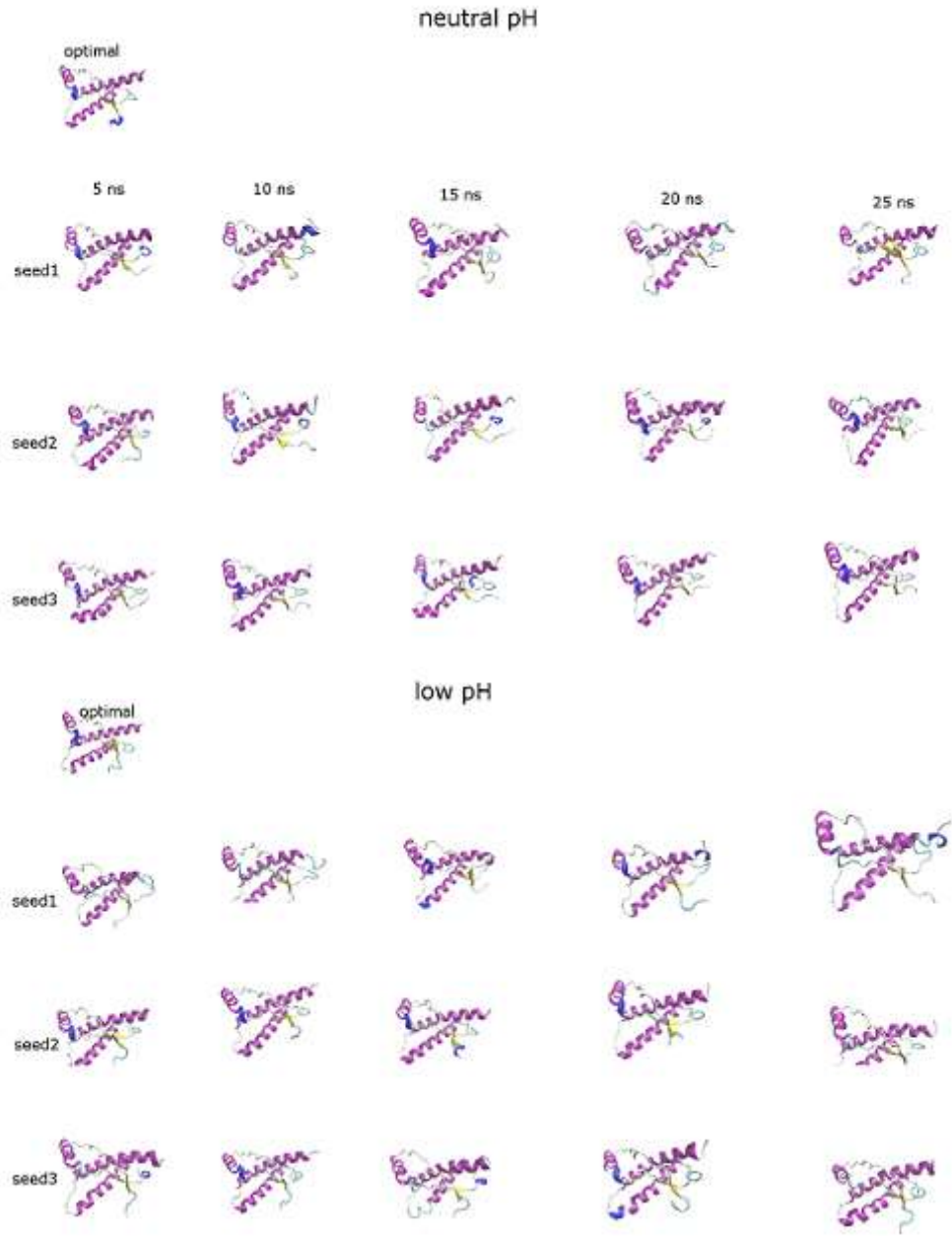

*Figure 11: Snapshots at minimized/optimal structure, 5 ns, 10 ns, 15 ns, 20 ns, and 25 ns (columns from left to right) for MD of homology structure of BufPrP$^C$(124-227) at 300 K. The first four rows are for neutral-pH environment and the last four rows are for low-pH environment. The second and sixth rows are for seed1, the third and seventh rows are for seed2, and the fourth and eighth rows are for seed3.*

Lastly, we observed the electrostatic potential surface charge distributions of BufPrP$^C$(124-227). For rabbit PrP$^C$, it carries a continuous area of positive charges on the surface (Wen et al., 2010b) (mainly constructed by residues HIS139, HIS176, ARG150, LYS203, and ARG147 in Figure 12), which is distinguished from other PrP$^C$. Observing Figure 12, we may see around the β2-α2 loop buffalo does not have a large land of continuous area of positive charges as rabbit. For rabbit, there are two positively charged residues ARG163 and ARG227 linking the β2-α2 loop and the C-terminal end of H3. FirstGlance in Jmol (bioinformatics.org/firstglance/fgij/) was used to detect all the charges of BufPrP$^C$(124--227): 10+ (7 Arg, 3 Lys) (4 His) and 14- (5 Asp, 8 Glu, 1 C-termini), in defining the following SBs under the neutral-pH environment: ASP178.OD1-ARG164.NH1 (4.6319 Å), GLU186.OE2-LYS62.NZ (3.3250 Å), GLU200.OE2-LYS204.NZ (3.5305 Å), ASP202.OD2/1-

ARG156.NH2/NE (3.4993 Å), and GLU186.OE1/2-ARG156.NH1/2 ( 4.4191 Å). We used Maestro 10.1 2015--1 (Academic use only) free package to draw the Poisson-Boltzmann electrostatic potential surface charges of our energy minimized/optimal structure (also confirmed by Swiss-PdbViewer 4.1.0) and the average structures of 25 ns' MD of homology BufPrP$^C$ at 300 K in neutral-pH environment (Figure 13). In Figure 13 (also in Figure 12), we can observe that the positively charged surface (blue coloured) at ARG164 in the β2-α2 loop, and a large continuous positively charged surface constructed by residues ARG136, ARG151, HIS140, ARG208, and LYS204 together. For BufPrP$^C$(124-227), LYS194, ARG156, or ARG148 constructs discrete area of the positively charged electrostatic surface respectively.

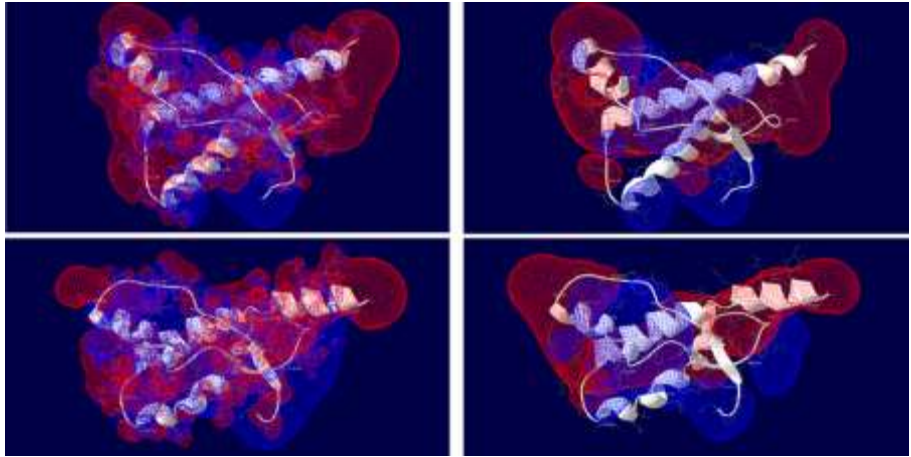

*Figure 12: Surface electrostatic charge distributions of the energy minimized/optimal structures of BufPrP$^C$(124-227) and rabbit PrP$^C$(124-228) (2FJ3.pdb). The first row is for buffalo and the second row is for rabbits. The first column is for all the atoms partial charged and the second column is for all the residues charged. Poisson-Boltzmann method is used. Blue: positive charge, Red: negative charge. This Figure was drawn by Swiss-PdbViewer 4.1.0.*

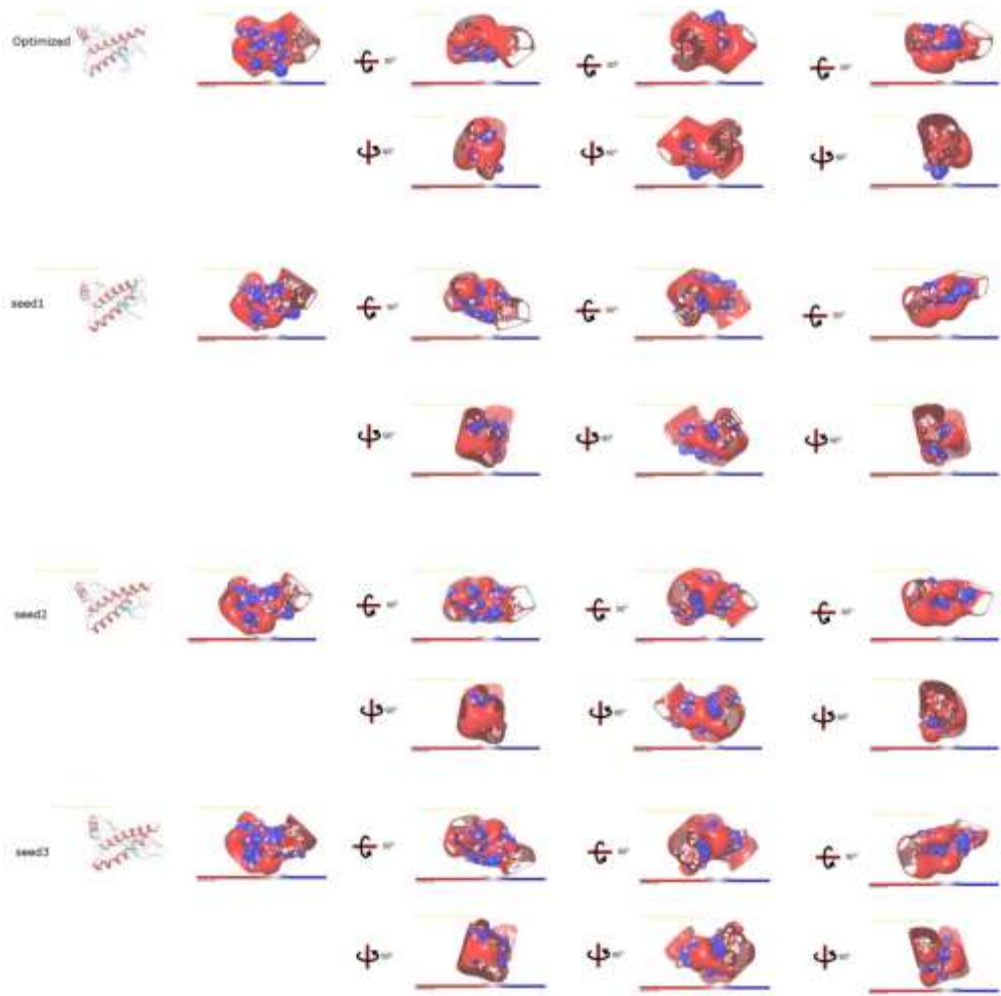

*Figure 13: Surface electrostatic charge distributions of the energy minimized / optimal structure and the average structures of 25 ns' MD of homology BufPrP$^C$ at 300 K in neutral-pH environment, where blue is for positive charge whereas red is for negative charge. Up to down: optimized structure, seed1-seed3. The pb_potential_volumes are ±29.3175, ±42.4453, ±42.499, and ±51.4525, respectively, for the optimized and seed1-seed3. Blue: positive charge, Red: negative charge. This Figure was drawn by Maestro 10.1 2015-1 (academic use only).*

## 4 Concluding Remarks

This study constructed a molecular structure of buffalo prion protein and then did MD study on this molecular structure. Clearly, this homology structure is useful as a reference for biochemical laboratories and later for NMR or X-ray structural laboratories. Buffalo is a low susceptibility species resisting prion diseases, and buffalo prion protein is very stable. To date, there is no structural data available. Protein structure of buffalo PrP was constructed by this study and we also present structural bioinformatics about molecular dynamics of buffalo PrP protein. Same as that of rabbits, dogs, or horses, the SB ASP178–ARG164 (O–N) (keeping the β2-α2 loop linked) contributes to the stability of buffalo prion protein. We also found HB SER170–TYR218 (linking the β2-α2 loop with the C-terminal end of α-helix H3) and SB HIS187–ARG156 (N–O) (linking α-helices H2 and H1) contribute to the stability of buffalo prion protein. At D178, there is a HB Y169–D178 and a polar contact R164–D178 for BufPrP$^C$ instead of a polar contact Q168–D178 for bovine PrP$^C$. Buffalo is a species with low susceptibility to prion diseases; thus, the bioinformatics of this study might be useful to the structure-based drug design of prion diseases. Rabbits, dogs and horses are also the species with low susceptibility to prion diseases; we found buffaloes, rabbits, dogs and horses all have a SB ASP178–ARG164 (O–N) keeping the β2-α2 loop linked. For buffalo prion protein, in the β2-α2 loop, there is a strong π-π stacking and a strong π-cation F175–Y169–R164.(N)NH2. The authors hope the bioinformatics presented in this study is helpful and useful for experimental studies of buffalo prion protein in laboratories.

**Acknowledgments:**

This research was supported by a Victorian Life Sciences Computation Initiative (VLSCI) grant number VR0296 on its Peak Computing Facility at the University of Melbourne, an initiative of the Victoria Government. We thank Chatterjee S for his contribution the beautiful Figure 1 to this paper.